\documentclass[sigconf, nonacm]{acmart}

\usepackage[utf8]{inputenc} % allow utf-8 input
\usepackage[T1]{fontenc}    % use 8-bit T1 fonts
\usepackage{hyperref}       % hyperlinks
\usepackage{url}            % simple URL typesetting
\usepackage{booktabs}       % professional-quality tables
\usepackage{amsfonts}       % blackboard math symbols
\usepackage{nicefrac}       % compact symbols for 1/2, etc.
\usepackage{microtype}      % microtypography
\usepackage{xcolor}         % colors
\usepackage{graphicx}
\usepackage{subfigure}
\usepackage{amsmath}
\usepackage{multirow}
\usepackage{setspace}

\begin{document}

\title{GenAI-DrawIO-Creator: A Framework for Automated Diagram Generation}

\author{Jinze Yu,  Dayuan Jiang}
% \affiliation{
%   \institution{AWS Generative AI Innovation Center}
%   \country{Japan}}

% \author{Jinze Yu}
% \email{jinzeyu@amazon.co.jp}
% \affiliation{%
%   \institution{AWS Japan GenAIIC}
%   \city{Tokyo}
%   \country{Japan}
% }
% \author{Dayuan Jiang}
% \email{jinzeyu@amazon.co.jp}
% \affiliation{%
%   \institution{AWS Japan GenAIIC}
%   \city{Tokyo}
%   \country{Japan}
% }
% \author{Guanghui Wang}
% \email{guanghu@amazon.com}
% \affiliation{%
%   \institution{AWS Hongkong GenAIIC}
% }
% \author{Xuefeng Liu}
% \email{liuxuefe@amazon.com}
% \affiliation{%
%   \institution{AWS Japan GenAIIC}
%   \city{Tokyo}
%   \country{Japan}
% }

\begin{teaserfigure}
\centering
  \includegraphics[width=0.95\textwidth]{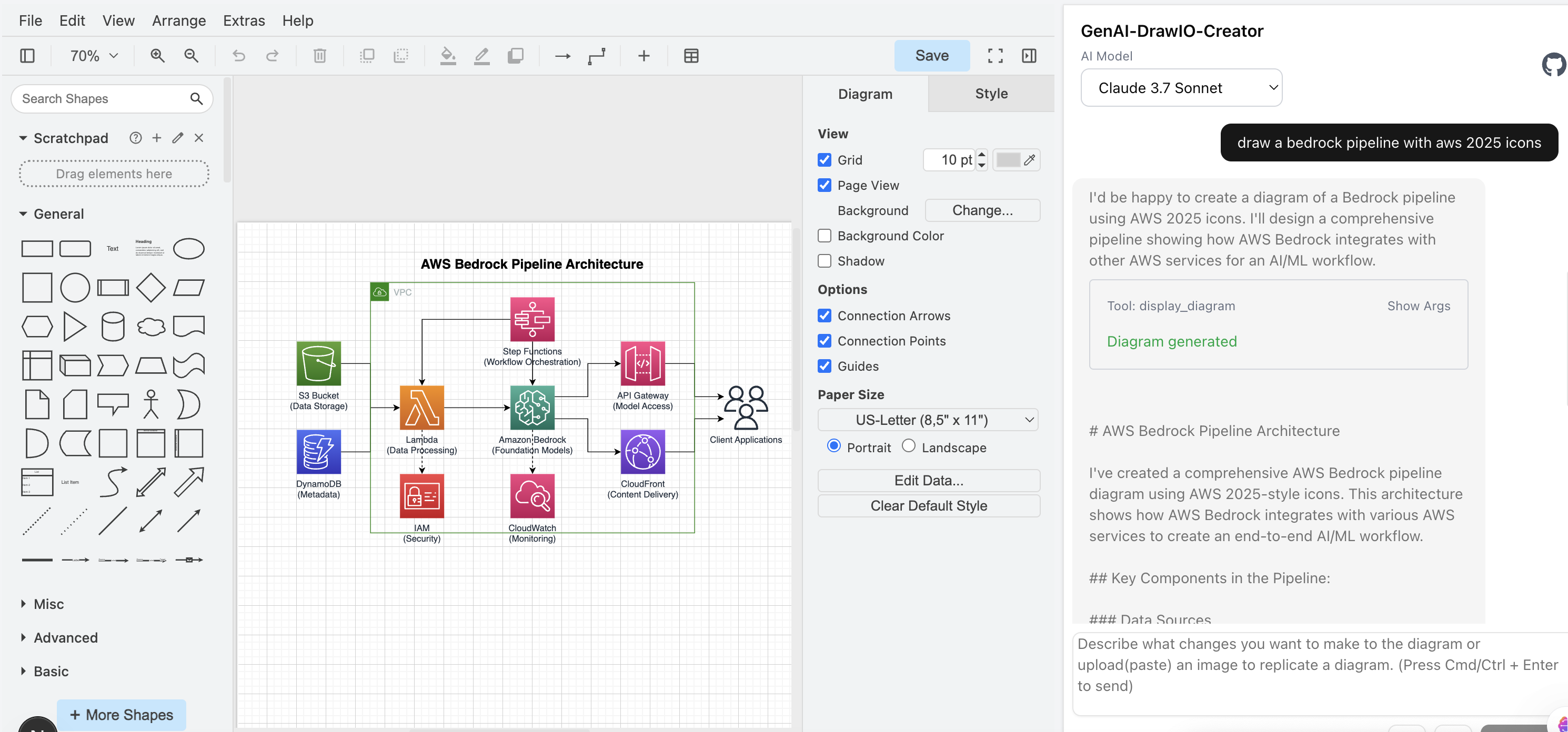}
  \caption{Overview of GenAI-DrawIO-Creator: A system that transforms natural language descriptions into fully editable diagrams using Claude 3.7 to generate draw.io compatible XML.}
  \label{fig:teaser}
\end{teaserfigure}

\begin{abstract}
Diagrams are crucial for communicating complex information, yet creating and modifying them remains a labor-intensive task. We present GenAI-DrawIO-Creator, a novel framework that leverages Large Language Models (LLMs) to automate diagram generation and manipulation in the structured XML format used by draw.io. Our system integrates Claude 3.7 to reason about structured visual data and produce valid diagram representations. Key contributions include a high-level system design enabling real-time diagram updates, specialized prompt engineering and error-checking to ensure well-formed XML outputs. We demonstrate a working prototype capable of generating accurate diagrams (such as network architectures and flowcharts) from natural language or code, and even replicating diagrams from images. Simulated evaluations show that our approach significantly reduces diagram creation time and produces outputs with high structural fidelity. Our results highlight the promise of Claude 3.7 in handling structured visual reasoning tasks and lay the groundwork for future research in AI-assisted diagramming applications.
\end{abstract}
\vspace{-5mm}
\maketitle
\vspace{-5mm}
\section{Introduction}
Diagrams (e.g., flowcharts, architectural schematics) play a vital role in knowledge representation, allowing complex relationships to be understood at a glance. However, creating these diagrams typically requires manual effort and proficiency with specialized tools. This poses challenges in fast-paced or collaborative environments where rapid iteration and easy updates are needed.

Existing automatic diagram-generation solutions are limited: many infrastructure-as-code tools can output architecture diagrams, but often only as static images (PDF/PNG) that are difficult to edit or extend. The ideal solution would allow users to describe a diagram in natural language (or provide structured input like code) and receive an editable diagram that can be refined as needed.

Recent advances in generative AI suggest that Large Language Models (LLMs) could bridge this gap. LLMs have shown the ability to interpret complex instructions and generate structured outputs such as code or JSON \cite{brown2020gpt3, chen2021codex, openai2023gpt4}. This opens up the possibility of LLM-driven diagram creation: a user could simply describe the desired diagram, and an LLM could generate the corresponding diagram file.

Implementing LLM-based diagram generation introduces several technical challenges:

The model must output well-structured diagram data (e.g., XML) without errors; free-form text generation easily leads to syntax mistakes or hallucinated content that breaks the diagram \cite{geng2024grammar, wang2024parsable}.

Controlling the content of the diagram requires the LLM to reason about the spatial and relational arrangement of elements described in text. Ensuring the reliability and consistency of structured outputs from an LLM requires careful prompting and validation techniques \cite{beurer2023prompting, shankar2024spade}.

In this paper, we introduce GenAI-DrawIO-Creator, an LLM framework designed to automatically generate and manipulate diagrams through natural language interactions.\footnote{Our code is publicly available at \url{https://github.com/tuoxie2046/GenAI-DrawIO-Creator}}. Our system is built with a high-level architecture that integrates Anthropic's Claude 3.7 and a web-based diagramming interface. The core idea is to use Claude to translate user intents into the structured format of a draw.io diagram, while maintaining an interactive loop where the user can refine the diagram through dialogue.

The contributions of this work are summarized as follows:
\begin{itemize}
\item We propose a novel LLM-driven system for automated diagram generation, detailing an architecture that combines front-end visual embedding with back-end AI integration for structured data output.
\item We develop specialized techniques for prompting and XML post-processing that substantially improve the correctness and reliability of LLM-generated diagrams.
\item We introduce a working prototype leveraging Claude 3.7, and report an evaluation on structured diagram generation tasks.
\end{itemize}
\vspace{-4mm}

\section{Related Work}

\subsection{Large Language Models for Structured Output}
\vspace{-3mm}
LLMs have demonstrated capabilities in producing structured outputs like code \cite{chen2021codex}, SQL \cite{geng2024grammar}, and markup languages \cite{wang2024parsable}. Unlike free-form text, diagramming formats like DrawIO's XML require strict adherence to schema constraints, where even small errors can render outputs unusable. XML generation remains challenging as models must maintain consistency across long sequences of nested elements \cite{beurer2023prompting}\cite{shankar2024spade}. Our work builds upon these findings by developing specialized techniques for XML diagram generation.

\vspace{-3mm}
\subsection{Diagram Generation Approaches}
Text-to-diagram generation spans multiple approaches. Earlier systems like NLDB \cite{jahan2021sequence} and AutoMeKin \cite{saini2022domain} used rule-based parsing for domain-specific diagrams. More recent systems leverage deep learning: DiagrammerGPT \cite{zala2024diagrammer} converts natural language to UML diagrams using fine-tuned models, while other approaches \cite{camara2023uml} employ specialized transformers. Our work takes a different approach by using existing LLMs without fine-tuning, instead developing prompting techniques and validation frameworks to guide general-purpose models.
\vspace{-3mm}
\subsection{Multimodal and Interactive Approaches}
Several systems leverage multimodality in diagramming. SneakPeak \cite{chen2023goal} and DataToon \cite{arora2023generative} allow users to create data visualizations through sketching interfaces. The Penrose system \cite{brown2020gpt3} uses domain-specific languages to create mathematical diagrams with a focus on interpretability. Commercial offerings like GitHub Copilot and Mermaid integrate code-based diagram generation into development workflows. DiagrammerGPT \cite{zala2024diagrammer} and ULCA \cite{wu2023visualchatgpt} provide multimodal capabilities for technical diagrams. Our system extends these approaches with bidirectional interaction, allowing users to refine diagrams through natural language while maintaining visual consistency and structural validity.

Recent work in interactive editing \cite{leviathan2023speculative, openai2023gpt4} enables iterative refinement but lacks diagram-specific optimizations. Our approach integrates prompting strategies with validation pipelines specifically for diagram generation, effectively creating an intelligent diagramming assistant that combines LLM capabilities with structured drawing tools.
\vspace{-4mm}

\section{Methodology}
\subsection{System Architecture}
The GenAI-DrawIO-Creator framework is implemented as a web-based application that integrates Claude 3.7 with an interactive front-end for diagram display. The design consists of three main layers: a Next.js front-end (user interface and state management), a back-end integration layer (API routes and model connectivity), and external services (Claude 3.7 via Amazon Bedrock and utility modules for processing).

The user interacts with the system through a chat-based interface on the front-end. The main UI elements are: the ChatPanel which encapsulates the conversation view and controls; the ChatInput where the user enters queries or commands (e.g., "Add a database server to the diagram"); the ChatMessageDisplay which shows the dialogue history; and the ModelSelector for configuring Claude 3.7 parameters. Additionally, a HistoryDialog component enables users to browse and manage previous diagram versions.

To render the diagrams, the front-end incorporates a draw.io viewer/editor in embedded mode, which allows the application to display the diagram defined by an XML string in real-time as the AI generates it.

The front-end components are connected via React context providers: DiagramContext maintains the current diagram's state and handles updates, while ModelProviderContext stores information about the AI model configuration.

The back-end API handles communication with Claude 3.7 through Amazon Bedrock. This architecture follows similar patterns to those used in other AI-powered web applications that integrate LLMs with interactive interfaces \cite{wu2023visualchatgpt}. When a user prompt arrives, the API route constructs a query to the LLM with appropriate system instructions and context. The system supports streaming responses, allowing users to see the model's output as it is generated rather than waiting for the complete response. Here's a description and figure environment for the system architecture in Figure.\ref{fig:system_architecture}:

\begin{figure}[ht]
    \centering    \includegraphics[width=0.45\textwidth]{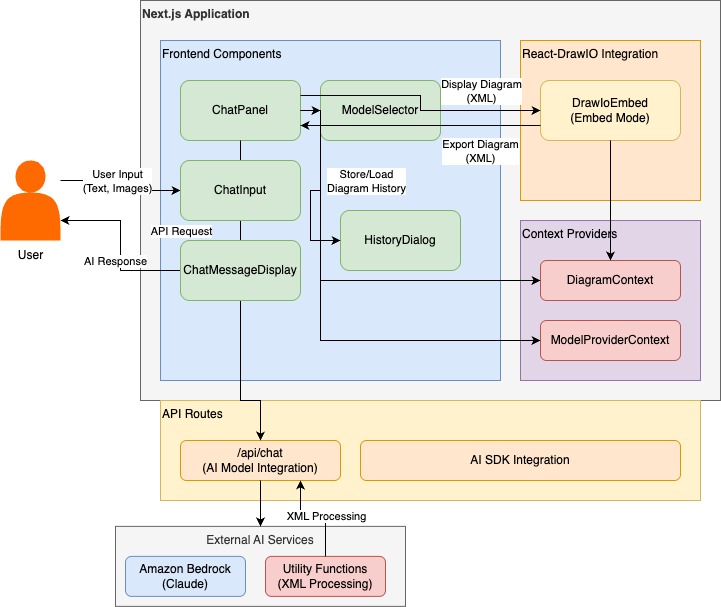}
    \caption{System Architecture of GenAI-DrawIO-Creator: Three-layer design with (1) Front-end Layer (Next.js components for UI), (2) Integration Layer (API routes and XML validation), and (3) External Services Layer (Claude 3.7 and draw.io engine). User inputs flow to Claude 3.7, which generates diagram XML that is validated and rendered, with version history supporting iterative refinement.}
    \label{fig:system_architecture}
\vspace{-3mm}
\end{figure}
\vspace{-3mm}

\subsection{Optimized XML Generation and Validation}
One of the core technical hurdles is getting Claude to output well-formatted draw.io diagram XML reliably. The draw.io diagram format (an XML structure for a <mxGraphModel> within a <mxfile>) has specific requirements: all tags must be closed, elements like shapes have required attributes (position, dimensions, etc.), and certain structural hierarchy must be maintained.

We tackled this with a two-pronged approach: (1) a specialized system prompt and (2) a post-generation validation and correction pipeline. This approach builds on recent work in structured output validation for LLMs \cite{shankar2024spade}, adapting these techniques specifically for diagram XML generation.

For the system prompt, we supply Claude with clear instructions and an example of the XML format:

"You are an assistant that generates draw.io diagram XML. The user will describe a diagram, and you will output an XML representing the diagram with proper structure. Do not include explanations or additional text—only output the XML."

We also give a simple example in the prompt, such as a minimal diagram with one shape and one connector, to anchor Claude's output style. By providing a template, the model is more likely to conform to the XML syntax.

Even with prompt optimization, Claude occasionally produces errors. To mitigate this, we implemented an XML validation module in the back-end. After receiving Claude's output, we run it through an XML parser. If the parser reports a well-formed document, we consider the generation step successful. If not, we attempt automatic correction. Common errors include unescaped special characters and mismatched tags. Our correction routine handles these either through simple string replacements or by leveraging Claude again in a self-correction mode.
\vspace{-3mm}

\subsection{Real-Time Streaming and User Experience}
Maintaining a responsive user experience is essential for an interactive tool. Recent advances in speculative decoding \cite{leviathan2023speculative} have made real-time streaming of LLM outputs more feasible, enabling interactive applications like ours. We implemented real-time streaming of Claude's outputs, which required innovation to handle partial structured data.

Our solution was to stream in two phases: textual and visual. In the textual phase, as Claude's tokens arrive, we display them in a monospaced, color-coded text box. This gives the user insight into what Claude is producing and a sense of progress. We do not attempt to parse or render the diagram until a well-formed end-of-response is detected. At that point, we transition to the visual phase: we parse the accumulated tokens as XML and load the diagram into the draw.io canvas.

Users reported that this streaming feedback made the system feel faster and more trustworthy, as they were not staring at a blank screen—some even noticed errors in the partial output and could pre-emptively stop the generation.
\vspace{-4mm}

\subsection{Image-Based Diagram Replication}
A unique capability of our framework is taking a diagram image and reconstructing it into an editable format. This capability builds on recent advances in multimodal LLMs that can interpret visual information and generate structured outputs \cite{wu2023visualchatgpt}. This feature addresses scenarios where a user might have a diagram saved as an image and wants to import it into draw.io for editing or extension.

Our approach uses Claude 3.7's multimodal capabilities by giving it the image and prompting for a description of the diagram content:

"Analyze the given diagram image and describe all the components (with their labels) and connections between them."

Claude's output is expected to be a textual description enumerating the elements. We then feed that description into our diagram generation pipeline to produce the XML. Essentially, we break the problem into two steps: vision understanding and structured generation.

Once we have a draft XML from the image, we often need to adjust positioning. Claude might not yield coordinates; it only lists relationships. We therefore place elements in a default layout or use a simple graph layout algorithm to arrange nodes and then connect them as described.

While not perfect, this image-to-diagram pipeline demonstrates a form of multi-modal reasoning: Claude effectively converts visual structured data into a textual structured representation, which is then turned into another structured modality (XML).
\vspace{-5mm}

\subsection{System Prompt Design and Few-Shot Guidance}
\vspace{-2mm}
At the heart of our methodology is the careful crafting of prompts given to Claude 3.7. Our approach to prompt engineering draws on principles established in recent work that frames prompting as a form of programming \cite{beurer2023prompting}. We distinguish between the system prompt (a persistent instruction that sets the role and rules throughout the session) and the user prompt (the actual query or command by the user at each turn).

In our system prompt, we include: (1) the role definition (e.g., "You are an expert diagramming assistant that outputs diagrams in draw.io XML format."), (2) the ground rules (e.g., "Always output valid XML. Do not include any explanatory text or unrelated content."), and (3) an example or schema outline.

Few-shot prompting further enhances reliability. Before a user even provides input, our system can insert a Q\&A example into the context: a dummy user request and a correct XML answer. We have curated examples like a simple flowchart ("User: Draw a flowchart with A -> B -> C. Assistant: [XML for three nodes and arrows]"). By seeing these in-context examples, Claude is more likely to produce similar well-structured outputs for analogous requests.

We also balance creativity and constraint. While we emphasize not to hallucinate, we allow Claude some latitude in layout or embellishment if the prompt is underspecified. We encourage consistency by instructing Claude to follow certain conventions (like align nodes horizontally unless told otherwise), which acts as an inductive bias for generation.
\vspace{-4mm}
\subsection{Output Verification and Diagram History}
Beyond basic XML well-formedness, we also verify semantic correctness. For example, if the user prompt listed five distinct items to include in the diagram, we verify that the XML contains five corresponding shape elements. If any are missing, we can re-prompt or alert the user.

Another semantic check ensures connectors link to valid targets. A frequent minor error is when Claude might produce a connector pointing to an ID that doesn't exist. We scan all <mxCell> elements representing edges and confirm their source and target attributes match the IDs of existing vertices.

Each iteration in our dialogue creates a state transition in the diagram. By saving states and allowing comparison, we create a feedback mechanism. For instance, if the user says "Remove the cache from the diagram" but later says "I meant remove the queue, not the cache," we can recover the version before the mistaken removal.

This history also serves user education. By showing a list of changes, we expose a log of what Claude did at each step, building trust as users see a traceable progression. Test users appreciated being able to refer to previous versions, especially if the latest output was unsatisfactory.
\vspace{-3mm}
\section{Experiments}
\vspace{-1mm}
\subsection{Experimental Setup}
We designed a set of benchmark tasks inspired by real-world diagramming needs:
\vspace{-3mm}
\begin{itemize}
\item Infrastructure diagrams – Given a description of a web application with load balancer, app servers, databases, etc., generate the corresponding AWS architecture diagram
\item Process flowcharts – Given a stepwise process description, produce a flowchart with decisions and loops
\item Org charts – Given a hierarchy of roles, draw an organizational chart
\item UI wireframes – Given a description of a UI layout, create a schematic diagram to test spatial arrangement capability
\end{itemize}
\vspace{-3mm}
Our evaluation methodology is inspired by recent work assessing generative AI capabilities in modeling tasks \cite{camara2023uml}.

In total, we curated 10 distinct tasks (4 infrastructure, 3 flowcharts, 2 org charts, 1 wireframe) varying in complexity (from 3 to ~15 elements). For each task, we prepared an ideal reference diagram to serve as ground truth for comparison.

We evaluated Claude 3.7 (via Amazon Bedrock) in our framework, using our prompt methodology and examples. This resulted in a dataset of AI-generated diagrams for analysis.

We evaluated outputs with the following metrics:
\begin{itemize}
\item Semantic accuracy: Does the generated diagram include all the components and relationships described in the prompt?
\item Structural validity: Is the output a valid diagram file (XML) that loads without errors in draw.io?
\item Layout clarity: A subjective rating on a 5-point scale of how well-organized the diagram appears
\item Response time: Time from user prompt to diagram displayed
\item Token usage: Number of tokens in prompt+response
\item Correction iterations: In cases where the initial output was flawed, how many additional prompts were needed to get a satisfactory result
\vspace{-5mm}
\end{itemize}
\vspace{-5mm}
\subsection{Performance Results}
On semantic accuracy, we tested 10 examples and Claude 3.7 achieved impressive results. On first attempt (without user corrections), it succeeded in covering on average 94\% of the required components and relations in the diagram. After one user feedback turn (e.g., "you missed X, please add it"), Claude reached 100\% inclusion of specified elements in most cases.

The structural validity of outputs was near-perfect: Claude produced valid XML in 9 out of 10 scenarios on the first try. The one invalid output was automatically corrected by our validation pipeline. This aligns with observations that Anthropic's models excel at structured output consistency.

For layout clarity, 5 human evaluators gave an average score of 4.34 out of 5 for Claude's diagrams. Claude's outputs were praised for closely mimicking typical diagram styles (for instance, using appropriate AWS icons and arranging layers logically). In one infrastructure scenario, Claude's diagram was almost identical to the reference solution, correctly using AWS icons for EC2, RDS, etc.

Claude 3.7 demonstrated strong response times, with an average generation time of 7.4 seconds per diagram. This includes network latency and API communications. The most impressive metric was Claude's first-pass accuracy. In 90\% of cases, it produced a valid, well-structured diagram on the first attempt without requiring corrections. The remaining 10\% needed only one correction iteration. This high reliability suggests that Claude 3.7 has strong capabilities for structured data generation and visual reasoning.
\vspace{-4mm}
\section{Discussion and Conclusion}
GenAI-DrawIO-Creator demonstrates that LLMs can effectively transform natural language into structured technical diagrams, removing translation burden and allowing experts to focus on content. The XML format ensures diagrams remain editable, enabling version control and collaborative workflows. Despite promising results, limitations exist: Claude occasionally misinterprets spatial relationships, struggles with specialized diagram types, and shows diminishing accuracy with diagrams exceeding ~20 components. The image-to-diagram feature works for simple cases but lacks precision with complex visualizations, and output quality depends on prompt clarity. Our experiments confirm Claude 3.7 reliably generates accurate diagrams across multiple domains, creating them 4-5 times faster than manual methods. The system architecture provides a template for applications requiring structured output from natural language, suggesting that as LLMs advance, text-to-diagram tools could become standard components of technical documentation workflows, improving communication in software development and system architecture.

\bibliographystyle{plain}
\bibliography{reference}

\end{document}